\documentstyle[12pt]{article}

\def \be  {\begin{equation}}
\def \ee  {\end{equation}}
\def \beq  {\begin{equation}}
\def \eeq  {\end{equation}}
\def \ba  {\begin{eqnarray}}
\def \ea  {\end{eqnarray}}
\def \baa {\begin{eqnarray*}}
\def \eaa {\end{eqnarray*}}
\def \lab #1 {\label{#1}}
\newcommand\bqa {\begin{eqnarray}}
\newcommand\eqa {\end{eqnarray}}

\newcommand{\bear}{\begin{array}}
\newcommand{\enar}{\end{array}}

\font\cmss=cmss12 
\def\inbar{\,\vrule height1.5ex width.4pt depth0pt}
\def\IC{\relax\hbox{$\inbar\kern-.3em{\rm C}$}}
\def\IZ{\relax{\hbox{\cmss Z\kern-.4em Z}}}
\def\IR{{\hbox{{\rm I}\kern-.2em\hbox{\rm R}}}}

\def\IP{{\hbox{{\rm I}\kern-.2em\hbox{\rm P}}}}
\def\II{\hbox{{1}\kern-.25em\hbox{l}}}

\def\1{\hbox{{1}\kern-.25em\hbox{l}}}
\newcommand{\bit}[1]{\mbox{\boldmath$#1$}}
\newcommand{\ft}[2]{{\textstyle\frac{#1}{#2}}}

\def \e  {\mathop{\rm e}\nolimits}
\def \be  {\begin{equation}}
\def \ee  {\end{equation}}
\def \ba  {\begin{eqnarray}}
\def \ea  {\end{eqnarray}}
\def \IP {\mathbb{P}}
\newcommand{\as}{\ifmmode\alpha_{\rm s}\else{$\alpha_{\rm s}$}\fi}
\newcommand\re[1]{(\ref{#1})}

\def \tr {\mbox{tr}}
\newcommand \vev [1] {\langle{#1}\rangle}

\def\beqn{\begin{eqnarray}}
\def\eeqn{\end{eqnarray}}

\def \labeltest #1 {\label{#1}}
\def\beq{\begin{equation}}
\def\eeq{\end{equation}}

\begin{document}


\hfill\parbox{40mm}

\vspace{1mm}

\centerline{\large \bf  Gauge Theories as String Theories; the First Results}

\vspace{7mm}

\centerline{\bf   A. Gorsky}

\centerline{\it Institute of Theoretical and Experimental Physics}
\centerline{\it B. Cheremushkinskaya 25, Moscow, 117259, Russia}

\vspace{1cm}
\centerline{\bf Abstract}
The brief review of the duality between gauge theories and closed strings
propagating in the curved space is based on the lectures given at
ITEP Winter School-2005.

\vspace{5mm}


\vspace{2mm}

\section{Introduction}
The idea of gauge/string duality is one of the most profound
in the realm of fundamental interactions. It influenced a lot   both sides of the
correspondence since the first papers \cite{maldacena,gubser,witten1}
and several new important ideas and results have emerged
in string and gauge theories. In these lectures we shall try to
provide a first approximation to the subject without  entering
into any serious technical details. Instead we shall try to explain
the main ideas behind and present examples of the results and
predictions which can be derived and which have been already derived
along this line of considerations. The literature on the subject
is enormous so it is really impossible to mention all important
contributions therefore we shall assume that all relevant
references can be found in the reviews which we shall mention.

According to the most optimistic viewpoint string theory is the
"theory of everything" however the essence of gauge/string duality
is more moderate - it relates the closed string theory on curved
manifolds and gauge theories with some amount of supersymmetry.
The origin of the duality  lies on the stringy side,
actually it can be considered as a kind of the duality between
open string which involves gauge boson as the massless mode and
closed string whose massless mode is identified with  graviton.
However open/closed duality is not enough at all to formulate
gauge/string duality explicitly and  more concepts are
necessary which are borrowed from the results derived during the last decade.
The essential ingredients of the picture  are the solitonic objects
in the string theory - D-branes which provide their worldvolumes for
the four-dimensional theories to live on. Since open strings can
end on D-brane \cite{polchinski1} the massless mode of the open string with both ends
on the D-brane yields the abelian gauge field on D brane
worldvolume. Moreover the theory on the stack of N D branes turns
out to be nonabelian U(N) gauge theory \cite{wittenbr}. Therefore we have a kind
of building blocks which can be used to generate the gauge theory which
we would like to work with.

Turning to the closed string side we recall that any massive
object influences the metric around and D-brane being the object
with finite tension is not the exclusion. In the most popular
version the superstring enjoys
ten dimensional target space and propagates
in the metric induced by  D branes. It is convenient to work with
the stack of N branes at large N which amounts to the simplified
version of the metric we shall mainly discuss in our lectures.
It turns out that the closed string actually feels the metric
produced by branes and the induced higher-form field as well since D-brane
is "charged" object with respect to the higher form field.
The very idea is to "forget" about the D-branes at all and take as
their "trace" the deformation of the metric and fluxes they have
created. The situation is somewhat similar to the familiar black
hole physics when the massive object is hidden inside the horizon of
the black hole. Actually the geometry produced by large N branes
with additional fluxes amounted from their charges provides the
complicated background where the closed string lives in. We shall
explain that it is the derivation of this background is the
essential and rather involved part of the whole story.

Let us make a few  historical remarks. The idea of the gauge/string
duality has been forwarded by Polyakov long time ago and developed
by him over a few decades. An important step was the understanding of the
role of the
"additional" coordinate (supplementing the four usual ones)  as a
renormalization-group scale \cite{polyakov97}.
The idea was further promoted by Klebanov \cite{klebanov97}
who demonstrated  the possibility of self-consistent account  of the
back reaction of the
gravity on the D branes and vise versa.

The developments culminated  in
the work of Maldacena \cite{maldacena} who proposed the duality
between  the gauge theory with the largest N=4
supersymmetry and closed string
propagating in   $AdS_5\times S^5$ geometry and in an external 4-form field
of constant strength. In the original Maldacena's formulation
importance was assigned to the massless string modes that is
string theory is reduced to the supergravity.
Also, the holographic principle was implicitly assumed
\cite{holo} since the gauge theory was effectively formulated at
the boundary of
$AdS_5$ and supergravity in the entire ten-dimensional space.

Just after the  Maldacena,s proposal it was shown that the
classical action calculated on the supergravity solution to the equation of
motion with the fixed boundary condition serves as the generating
function for the correlators in the boundary N=4 gauge theory \cite{witten1,gubser}.
Later the corresponding backgrounds have been found for the gauge
theories with less amount of supersymmetry \cite{mn,ks,polchinski20}.
Moreover the
example of the geometry was described in which the string
theory is exactly solvable ; its detailed comparison with the
corresponding sector of the gauge theory confirmed the validity
of the dual description \cite{bmn}.

Our lectures are organized as follows. First we shall introduce
some notations and  formulate the Maldacena's conjecture for N=4
supersymmetric  Yang-Mills (SYM) gauge theory. In Section 3 we shall consider the duality
for N=4 theory in the supergravity approximation.
We briefly explain how the correlators in the gauge theory can
be calculated via the supergravity solution. As
the example of the remarkable prediction of the duality for the N=4
theory in the strong coupling hydrodynamical regime we briefly
explain the derivation of the shear viscosity.
In Section 4  we turn to the approximation
of the classical string as an example of duality in N=4
theory and  explain the calculation of the circular Wilson
loop on the both sides of the correspondence.

Section 5 will be devoted to  more detailed comparison
and prediction of the duality to
the matrix of the anomalous dimensions
of the gauge theory operators in SUSY theories. We shall
explain how the loop counting can be followed on the stringy side
and emphasize the important role played by the hidden integrable
structure reflecting the hidden symmetries of the problem. The analysis
in this section is based mainly on the classical string approximation.

The
example when the exact quantum spectrum can be found explicitly
shall be briefly discuss in the limiting geometry of the pp-wave
in Section 6. In
this case the mapping of the stringy states with the gauge theory
operators is very illuminating.
Then we shall reduce
SUSY on the gauge side down to N=2 and explain how the features of
N=2 theory manifest themselves on the supergravity side. In the
next section similar analysis will be done for N=1 theories which
have a lot in common with the realistic QCD case. In particular we
explain how the condensates, exact $\beta$ function  e.t.c can be
derived in the supergravity approximation on the stringy part. A few
results concerning the YM theory without supersymmetry will
be considered in Section 9.

The literature on the duality between gauge
and string theories is abundant hence we address the reader to the reviews
to learn the background material. The general review on
the correspondence can be found in \cite{obzor,klebanov20} the review
on the Wilson loop calculations is  \cite{sz}. Various
aspects of the hidden integrability in the  context of
gauge/string duality are presented in \cite{tsey1,bbgk,beisert}.
The  supergravity approach to N=2 theories is reviewed in  \cite{polchinski20}
and to N=1 theories in \cite{bertolini1,strassler}. The exactly solvable
string limit in the geometry of pp-wave and its relation
to the special operators in the N=4 gauge theory can be found in \cite{ppobzor}.

\section{Basics of gauge/string duality}

The duality is initially formulated for maximally supersymmetric
four-dimensional conformal N=4 gauge theory with vanishing $\beta$-
function. The fields of the theory include vector fields
four fermions and six real scalar fields $\Phi_i$
in the adjoint representation of the gauge group. There is global
SO(6) symmetry corresponding to the rotation of scalars and
fermions. Theory has nontrivial vacuum manifold of zero energy states parameterized
by the vacuum expectation values of the scalars. The action of the theory
can be written as
\beq
S_{N=4} =\frac{1}{g_{YM}^2}\int d^4xTr[F_{\mu\nu}^2 +(D\Phi_i)^2
+[\Phi_i,\Phi_j]^2] + fermions
\eeq

It is conjectured to be dual to the type IIB superstring
in $AdS_5\times S^5$ background \cite{maldacena}.
The
background metric in the Poincar\'e coordinates looks as follows
\begin{equation}
ds^2
=
\frac{r^2}{R^2}(-dt^2 +dx_1^2 +dx_2^2 +dx_3^2 )
+
R^2\frac{dr^2}{r^2}
+
R^2d\Omega^{2}_{5}
\end{equation}
where r is the radial coordinate in $AdS_5$ and the last term represents
$S^5$ part of the geometry.
The metric can be considered as the near horizon limit of the D3 brane metric. Since the
D3 brane is the source of the Ramond-Ramond four form field $A_4$ the background
solution is supplemented by the flux of the corresponding field strength
\begin{equation}
F_5 = d A_4,\qquad 
\int_{S^5}*F_5=N_c\,.
\end{equation}
The radii of the $AdS_5$ and $S^5$ are identical and equal to
\begin{equation}
R^4= 4\pi g_s \alpha^{\prime 2} N_c\,.
\end{equation}
where $g_s$ is the string coupling.
The four-dimensional gauge theory is localized on the boundary of the $AdS_5$.
The conformal $SO(2,4)$ group and the $R-$symmetry $SO(6)$ group of the
${N}=4$ SYM theory are identified with the isometry group of the $AdS_5$
and $S^5$ spaces, respectively. According to the gauge/string duality, the
eigenvalues of the dilatation operator in the ${N}=4$ SYM theory and the
energy spectrum of the string in the radial quantization coincide.
The
additional six coordinates  on the gauge theory side can be
identified with the vacuum eigenvalues of the   three complex
scalars $\Phi_i$ which belong to N=4 supermultiplet.

The next question must be the mapping of the parameters
in the dual theories. On the
gauge theory side we have coupling constant $g_{YM}$ and the rank
of the gauge group $N_c$. On the stringy side there are radii of the
background geometry $R_{S^5},R_{AdS_5}$ and the string coupling constant
$g_s$. The effective dimensionless tension of the string is expressed through the radius
as $T=\frac{R^2}{2\pi\alpha^{'}}$. The relation between parameters
reads as

\beq
4\pi g_s=g_{YM}^2,\qquad T=\frac{1}{2\pi}\sqrt{g_{YM}^2N_c}=
\frac{1}{2\pi}\sqrt{\lambda}
\eeq
which means that generically the quantum string in the curved
background with RR flux is involved into the correspondence. This
is very difficult object to work with however several
simplified limits
can be treated more successfully. First, it is possible to
consider the  limit of non-interacting  strings $g_s\to 0$
with
fixed tension T which to some extend corresponds to the classical
string.  If we also assume that
$T\to \infty$ the system reduces to supergravity approximation
involving only massless stringy modes.

In these lectures we shall consider examples when one or another limit
turns out to be fruitful. Several strong coupling phenomena shall
be manifest in the supergravity approximation. Some calculations can
be done at the level of the classical string theory including the
Wilson line calculation and anomalous dimensions. We shall also briefly
discuss the example of the pp-wave limit when the exact quantum stringy spectrum can be
found and compared with Yang-Mills results.

\section{Supergravity approximation}

 Going further one can assume that t'Hooft coupling is large
that is $T\to \infty$ which implies that only supergravity
massless modes survive on the stringy side and all massive stringy
modes decouple. It is this form of the correspondence between
strong t'Hooft coupling regime at the gauge side and the
supergravity limit on the stringy side which was considered by Maldacena.
The stringy sigma model in this limit is substituted by
the massless modes with IIB supergravity classical  action
\ba
&&
Z(G_{\mu\nu},B_2,\Phi,A,C_{2},A_{4})= \int d^{10} x \sqrt{-detG} e^{-2\Phi}
[R+4G^{\mu\nu}\partial_{\mu}\Phi\partial_{\nu}\Phi -\frac{1}{12}H_3^2] -
\nonumber\\
&&
\vspace{0.5cm}
-\frac{1}{2}\int d^{10} x \sqrt{-detG}[G^{\mu\nu}\partial_{\mu}A\partial_{\nu}A +
(F_{3}- AH_3)^2 +(F_{5}-\frac{1}{2} C_2 \wedge H_3 +
\frac{1}{2} B_2 \wedge F_3)^2]
\nonumber\\
&&
\vspace{0.5cm}
+\frac{1}{2} \int d^{10} x A_4 \wedge H_3 \wedge F_3 +
fermions
\ea
including the metric $G_{\mu\nu}$, two scalars $\Phi$ and $A$ , two
2-form fields $C_{\mu\nu}=C_2$ and $B_{\mu\nu}=B_2$ with field strengths $F_3$ and  $H_3$
and 4-form field $A_{\mu_1,\mu_2,\mu_3,\mu_4}=A_4$ with curvature $F_5$.
In what follows the condition of selfduality is imposed on the $F_5$.

The first check of the correspondence was done in \cite{gubser,witten1}
where the correlators on the gauge side were compared with the
solution to the classical equation of motion on the supergravity
side with the proper boundary condition. It was shown that
the supergravity action calculated on the classical solution with given
boundary data is a generating function for correlators in the gauge theory:

\beq <exp(\sum \phi_kO_k)>_{N=4}=exp(-S_{cl}^{sugra}(\phi_k(x,z)\to\phi_k(x)))
\eeq
where $O_k$ is  operator in the gauge theory  coupled to
the supergravity mode $\phi_k(x,z)$ which tends to  $\phi_k(x)$
on the boundary of $AdS_5$ space.

\subsection{N=4 SYM in supergravity approximation: calculation of correlators}

Let us discuss the simplest  example of the calculation when the duality can
be checked. To this aim consider dilaton field $\Phi$
in the $AdS_5$ background with the action  in the linear
approximation

\beq S(\Phi)=const\int d^4x dz\frac{1}{z^3}[(\partial_z\Phi)^2+
(\partial_m\Phi)^2] \eeq
where $m=1,\dots,4$ and metric

\beq
ds^2=\frac{R^2}{z^2}(dz^2+dx_m^2)
\eeq
with boundary at $z=0$ is assumed. The action diverges on the
classical solutions regular at the boundary and falling at large $z$
\cite{gubser,witten1} which requires the IR regularizing cutoff in
$AdS_5$ at $z=\epsilon$.

The proper normalizable solution to the equation of motion for the
dilaton field with the boundary condition $\Phi(z=\epsilon,
x)=e^{ikx}=\Phi_0(x)$ is

\beq \Phi(x_m,z)=\frac{(kz)^2K_{2}(kz)}
{(k\epsilon)^2K_{2}(k\epsilon)}e^{ik_mx_m}
\eeq
where $K_2$ is the modified Bessel function and $k=\sqrt{k_m^2}$.
The calculation yields for the action on this solution

\beq
S\propto N\int d^4x\int
d^4y\Phi_0(x)\Phi_0(y)\frac{1}{(\epsilon^2+ |x_m-y_m|^2)^4}
+O(\epsilon^2)
\eeq

On the gauge theory side dilaton field interacts with the
operator $trF^2$ and generating function is given by

\beq Z(\Phi_{0})= <exp(\frac{i}{g_{YM}^2} \int d^4x\Phi_0(x)tr[F^2+ \dots])>
\eeq
where dots stand for superpartners and averaging is performed
via the standard path integral for N=4 theory. At the quadratic
approximation in the dilaton field we obtain for the generating
function

\beq
Z(\Phi_0)\propto exp(- ai\int d^4xd^4y <trF^2(x)trF^2(y)>)
\eeq
where $a=const$.
Taking into account that conformal invariance fixes the correlators
\beq
<trF^2(x)trF^2(y)>\propto \frac{N^2}{|x_m-y_m|^8}
\eeq
we see that the classical solution in gravity indeed yields the
correct value of the correlator on the gauge theory side. The only
subtle point concerns the regularization. If we introduce the
parameter $\eta_{UV}$ to regularize the UV divergence at $x_m=y_m$
then expressions at gravity and gauge sides coincide explicitly if
$\eta_{UV}=\epsilon$. Similar checks have been done for many
correlators involving more complicated operators.

\subsection{N=4 SYM in supergravity approximation: shear viscosity}

Very interesting and rather unexpected application of the gauge/string
duality involves the calculation of the ratio of the shear viscosity to
the volume density of entropy in the hydrodynamical phase of the strongly
interacting gauge theory. It turns out that this ratio manifests the
universal features which can be captured from the dual gravity
description in the background of black hole in the AdS space.

To consider the thermal field theory we have to fix the corresponding
gravity background. The relevant part
of the background geometry  was identified in \cite{witten2} as the $AdS_5$ black
hole
\beq
ds^2=\frac{r^2}{R^2}[-(1-\frac{r_0^4}{r^4})dt^2 +dx^2 +dy^2+dz^2] +
\frac{R^2}{(1-\frac{r_0^4}{r^4})}dr^2
\eeq
The temperature in the field theory coincides with  the Hawking temperature
of the black hole moreover the entropy of the field theory is equal
to the entropy of the black brane, proportional to the area of the
event horizon
\beq
S=\frac{A}{4G}
\eeq
where G is Newton's constant. It is natural to consider the
volume density of the entropy dividing it by the infinite volume
factor along directions parallel to horizon.

The viscosity is calculated using the Kubo formula via
the equilibrium correlation function of the components of the
energy stress tensor
\beq
\eta=lim_{\omega\to0}\frac{1}{2\omega}\int dt d\vec{x}
<T_{xy}(t,\vec{x}),T_{xy}(0,0)> e^{i\omega t}
\eeq
where $T_{xy}$ are the components of the energy stress tensor
of the supersymmetric gauge theory.

Using the optical theorem the correlator can be related to the
absorbtion cross section of the graviton propagating
normally  to the brane and polarized in $xy$ plane \cite{klebanov97}
\beq
\eta=\frac{\sigma_{abs}(\omega=0)}{16\pi G}
\eeq
It was argued that
the absorbtion cross section for the graviton
coincides with the cross section for the scalar and
in the low-energy limit  equals to the area of the horizon. Hence
the following relation holds for the ratio \cite{kovtun}
\beq
\frac{\eta}{s}=\frac{\hbar}{4\pi k_{B}}
\eeq
Amazingly this result is universal. Moreover
it was demonstrated that it is  independent   on the
details of the metric in the supergravity background and  only
the presence of event horizon is essential.

There are some further arguments that the above result
set the lower bound for this ratio for any
relativistic field theory at finite temperature and vanishing
chemical potential. The first correction to this result
is in agreement with this  general conjecture \cite{corr}
however the status of this claim remains an open issue.

\section{N=4 SYM and classical string: Wilson loop}

When the tension of the string is large but finite
the approximation of the classical string is reasonable.
That is the problem is reduced to the solution to the
classical equation of motion with the proper boundary
conditions on the worldsheet. Classical $\sigma$ model
in $AdS_5 \times S^5$ solves the problems of the
strong coupling limit on the gauge theory side.

The interesting application of the classical string
case concerns the calculation of the circular Wilson
loop in N=4 theory which can be associated with the
worldline of heavy W-boson.
It is assumed that SU(N+1) gauge symmetry is broken down to
$SU(N)\times U(1)$ and the symmetry breaking condensate is large.
The corresponding phase
factor in the supersymmetric case involves both gauge field and scalars
\beq
W(C)=\frac{1}{N}trPexp[\oint d\tau(iA_{\mu}(x)\dot{x}^{\mu} +
\Phi_{i}(x)\theta^{i}|\dot{|x|})]
\eeq
where C is closed contour parameterized by $x^{\mu}(\tau)$,
$\theta_i$ is the unit vector in the direction
corresponding to the symmetry breaking. It turns
out that the calculation of the circular Wilson loop
can be done in the strong coupling limit of the gauge
theory and the result exactly fits the classical
string calculation.

The object which can be calculated in dual theories is
\beq
S=e^{-ML(C)}<W(C)>
\eeq
at the large mass M. On the stringy side one
calculates the path integral with worldsheet boundary coinciding with loop C
located at the boundary of $AdS_5$. Path integral  can be
written explicitly  as
\beq
S=\int
DX^{\mu}DY^{i}Dh_{ab}D\Theta^{\alpha}exp(-\frac{\lambda}{4\pi}
\int_{D}d^2\sigma\sqrt{h}\frac{h^{ab}}{Y^2}
(\partial_{a}X^{\mu}\partial_{b}X^{\mu}+\partial_{a}Y^{i}\partial_{b}Y^{i})
+fermions
\eeq
where $X$  and $Y$  are coordinates in the target manifold.

At large t'Hooft coupling the superstring path integral
can be treated semiclassically and calculated by the saddle point
approximation from bosonic Nambu-Goto action yielding
\cite{Maldacena98}
\beq
-log<W(C)>=\frac{\sqrt{\lambda}}{2\pi}Area(C) -ML(C)
\eeq
Area of
the worldsheet with boundary contour C diverges however the
divergence is proportional to the length of the boundary contour
and can be related to the mass of the W-boson. Therefore the finite
part of the area determines the Wilson loop expectation value and
the generic prediction of the gauge/string correspondence
at strong t'Hooft coupling is
\beq <W(C)>=exp(\sqrt{\lambda}c)
\eeq
where c is positive number depending on the contour. More
precisely the universal form reads as
\beq
<W(C)>=\lambda^{-3/4}e^{\frac{\sqrt{\lambda}}{2\pi}A(C)}\sum_{n=0}^{\infty}c_n\lambda^{-n/2}
\eeq

It is very instructive to consider the circular loop case. The
leading expression for the strong coupling regime from the
classical string is
\beq
<W(C)>=\sqrt{\frac{2}{\pi}}\lambda^{-3/4}e^{\sqrt{\lambda}}
\label{circular}
\eeq
It turns out that for the supersymmetric
circular loop the resummation of the perturbative series can be
performed in N=4 SYM theory \cite{esz}. The crucial fact which actually is
responsible for the existence of the exact answer is the complete cancellation of
the internal vertexes in higher loops in the perturbative series.
Therefore only planar diagrams matter and summation of all rainbow
diagrams results in the strong coupling limit exactly to
(\ref{circular}) \cite{esz}. Thus calculation of the Wilson loop
provides an  accurate test for duality in
the classical string approximation.

\section{N=4 SYM and classical string: hidden integrability and anomalous dimensions}

\subsection{General remarks on integrability in the gauge theories}
In this section we shall try to demonstrate that a hidden integrability
allows verifying duality in lowest orders of the perturbation theory.
Of primary importance is the identification of the anomalous dimensions
of the operators in the gauge theory with the energies of the string
configurations.

To start with it is necessary to explain the place of integrability in four
dimensional gauge theories. At first glance, the connection
between four-dimensional gauge theories and integrable systems may
seem surprising. Indeed, in the latter case one is dealing with
quantum-mechanical systems which have a finite number of degrees
of freedom and the same number of conserved charges. For such
systems the degrees of the freedom are just phase-space variables
involved in the Hamiltonian and the ``evolution time'' has a
literal physical meaning. In contrast, the Yang-Mills theories in
four dimensions are complex systems with infinitely many degrees
of freedom which are not integrable \textsl{per se}. Complete
integrability emerges as a unique feature of \textsl{effective}
Yang-Mills dynamics in various limits \cite{Lipatov94,FK95,Korchemsky95,BDM98}. The relevant degrees of
freedom, Hamiltonians and ``evolution times'' within the
Yang-Mills theory are different in different limits and their
identification is not a priori obvious.
It is the  scale dependence of composite (Wilson) operators in QCD and
SYM theory.
The dynamics described by an integrable model
corresponds to the scale dependence of corresponding observables with the
``evolution time'' being identified as a logarithm of the relevant energy
scale.
The integrable systems that emerge in this context turn out to be
related to the celebrated Heisenberg spin magnet. This
model  describes a one-dimensional
chain of spins with an exchange interaction,
\begin{equation}
\label{XXX}
{H}_{s=1/2}
=
- \sum_{n=1}^L \left({S}_n \cdot {S}_{n+1} - \frac14 \right)
\end{equation}
where ${S}_n = (S_n^x, S_n^y, S_n^z)$ is the spin $-1/2$ operator of
the $n-$th site in the chain of length $L$ and periodic boundary conditions
are implied ${S}_{L+1}={S}_1$. The model (\ref{XXX}) is
completely integrable and its eigenspectrum was been found by  the Bethe Ansatz.
Much later it was understood \cite{KRS81,TTF83} that the original Heisenberg
model (\ref{XXX}) can be generalized to arbitrary spins while preserving
complete integrability. The Hamiltonian of a completely integrable lattice
model describing a chain of interacting spin$-s$ operators was found to be
\cite{TTF83}
\begin{equation}
\label{XXX-gen}
{H}_{s}
=
\sum_{n=1}^L H(J_{n,n+1})
\, , \qquad
J_{n,n+1} (J_{n,n+1} + 1) = ({S}_n + {S}_{n+1})^2
\, .
\end{equation}
Here the operator $J_{n,n+1}$ is related to the sum of two spins in the
neighboring sites, ${S}_n^2 = s(s+1)$, and $H(x)$ is the following
harmonic sum
\begin{equation}
{H}(x) = \sum_{l=x}^{2s - 1} \frac1{l + 1} =  \psi(2 s + 1) - \psi(x + 1),
\label{XXX-gen1}
\end{equation}
where $\psi(x)=d\ln\Gamma(x)/dx$ is the Euler $\psi-$function. For $s=1/2$ the
two-particle spin takes the values $J_{n,n+1}=0$ and $J_{n,n+1}=1$. In that case,
$H(0)=1$ and $H(1)=0$ so that the two-particle Hamiltonian $H(J_{n,n+1})$ is
given by a projector onto $J_{n,n+1}=0$ subspace, $H(J_{n,n+1}) = \frac14 -
{S}_n \cdot {S}_{n+1}$, in agreement with (\ref{XXX}).

Approximately at the same time as the model (\ref{XXX-gen}) has been formulated,
QCD calculations of anomalous dimensions of twist-two Wilson operators
and high-energy asymptotics of scattering amplitudes \cite{BFKL} have lead to
expressions involving the very same combination $[\psi(J) - \psi(1)]$ with $J$
having the meaning of the Lorentz and conformal $SL(2)$ spin, respectively. In
both situations the appearance of the $\psi$-functions is a
generic feature related to the existence of massless vector fields (gluons). For
almost two decades this similarity remained unnoticed mainly because of virtually
no interaction between the two communities.
Matching the QCD expressions into (\ref{XXX-gen}) one
discovers the hidden integrability properties of gauge
theories~\cite{Lipatov94,FK95,Korchemsky95,BDM98}.

{}In the so-called  Bjorken kinematic limit for ``hard''
scattering processes involving a large momentum transfer to a hadronic system,
a short-distance perturbative QCD dynamics can often be separated (factorized)
from the large-distance interactions and described through a set of gauge
invariant local composite operators built from fundamental fields and covariant
derivatives. These (Wilson) operators mix under renormalization and their scale
dependence is governed by the renormalization group (RG) or Callan-Symanzik
equation
\begin{equation}
\label{RG}
\mu\frac{d}{d\mu} {O}_N (x) = \sum_K \gamma_{NK}(g) {O}_K (x)
\, ,
\end{equation}
where $\gamma_{NK}$ is the mixing matrix given by a series in the
running coupling constant $g = g (\mu^2)$. The size of the mixing
matrix is constrained by the symmetries and depends on the
operators under consideration. The matrix $\gamma_{NK}$ can be
interpreted as a Hamiltonian acting in the space of operators that
get mixed via the RG flow \cite{BFLK85} with logarithm of the RG
scale $\tau=\ln\mu$ playing the role of  the ``evolution time''.
In this way, the evolution equation (\ref{RG}) takes the form of a
Schr\"odinger equation. It turns out that for a certain subclass
of operators  and to one-loop accuracy the corresponding
Hamiltonian can be identified as that of the open and/or closed
Heisenberg magnet with the spins being the generators of the
$SL(2,{R})$ group \cite{BDM98,Belitsky00,DKM00}. The number
of sites in the spin chain is given by the number of fundamental
fields involved in the composite operators and the value of the
spin at each site is fixed by the $SL(2,{R})$
representation to which the corresponding field belongs to. It is
different for quarks and gluons. The emergence of
$SL(2,{R})$ group as a symmetry group of the spin chain is
not accidental since this group is just a reduction of the
four-dimensional conformal group $SO(2,4)$ for field operators
``living'' on the light-cone \cite{Makeenko:bh}.
Integrability allows one to apply the Bethe Ansatz to reconstruct
the spectrum of the anomalous dimensions
\cite{Korchemsky96,Belitsky00,DKM00}. The work in this direction
\cite{BDKM99,Belitsky99a,Belitsky99b,BKM00,BKM01} has lead to an
almost complete understanding of the spectrum of anomalous
dimensions of twist-three operators in QCD which are important for
phenomenology. It has to be mentioned that, as a rule, QCD
evolution equations only become integrable in the large $N_c$
limit. There is no chance that a quantum $SU(N_c)$ theory would
turn out integrable for any $N_c$ because the phenomena it
describes are too complicated.

Although historically the phenomenon of integrability has been first discovered
in QCD for operators with maximum helicity, it is in fact a general hidden
symmetry of all Yang-Mills theories which is not manifest on the classical level
and is enhanced in its supersymmetric extensions \cite{BS03}. In a supersymmetric
Yang-Mills theory the mixing matrix in (\ref{RG}) gets modified due to the
presence of additional fields. Their contribution preserves the QCD-type
integrability and further augments it to an increasingly growing class of
operators as one goes from pure Yang-Mills (${N}=0$) to the maximally
supersymmetric ${N}=4$ SYM theory \cite{BDKM04}. In particular, in
the ${N}=4$ SYM theory integrability was rediscovered in the sector of scalar
operators \cite{mz,BeiKriSta03}. This theory involves three complex scalars
and the one-loop mixing matrix for the scalar operators of the type ${\rm tr}\,
\{\Phi_1^{J_1}(0)\Phi_2^{J_2}(0)\Phi_3^{J_3}(0)\}$ can be identified as the
Heisenberg $SO(6)$ spin chain with $J_1+J_2+J_3$ sites. The $SO(6)$ group
is nothing else than the $R-$symmetry group of the model with six real scalars
belonging to its fundamental representation. As a natural generalization of
these two integrable structures, discovered independently, it was found that
the one-loop renormalization of operators involving gauge, fermions and scalar
fields in ${N} = 4$ SYM is described by a spin chain with $SU(2,2|4)$
group representations on each site \cite{BS03}.

The BMN operators  \cite{bmn} is the only well established example in which the exact quantum
answer on the stringy side can be matched with the all-loop anomalous dimension
on the gauge theory side. Later in this section we shall discuss generalized BMN
operators which can be treated semiclassically on the stringy side and
demonstrate that the corresponding solutions to the classical equations of motion
in the stringy sigma models are ultimately related to classical integrable
models. We shall explain the relation between the string sigma model and quantum
spin chains in the thermodynamical limit and demonstrate that the anomalous
dimensions of the certain operators are in the one-to-one correspondence with the
special class of classical solutions to the sigma model for which this model
reduces to finite dimensional integrable systems of the Neumann type. Finally, we
shall comment on the relation between general classical solutions to the sigma
model and the Bethe anzatz solution to the compact quantum spin chains in the
semiclassical limit.

\subsection{Derivation of the string in the thermodynamical limit of the spin chain}
The hidden integrability in YM theory involves Heisenberg spin
chain as one-loop dilatation operator. The number of cites in the
chain equals to the number of fields involved in the composite
operator, for instance operators of the type $Tr\Phi^N$ where
$\Phi$ is some field in the theory correspond to the chain of
length N. At large N one can consider thermodynamical limit of the
spin chain. It turns out that thermodynamical limit can be matched
with the  action of the string propagating in the submanifold of
$AdS_5\times S^5$ geometry that is spin chain in fact should be
considered as discretization of the string in curved background.

Let us consider example of such derivation and explain how  the
action of the string moving on $S_3$ can be obtained from
XXX SU(2) spin chains describing the renormalization of the YM
operators involving two scalars $\Phi_1,\Phi_2$. The sigma model
describing the string moving in the appropriate curved background
can be derived from the  SU(2) spin chain in the long wave length
limit. The corrections to the classical sigma model behave as
${1}/{J}$, where the angular momentum of the string $J$
corresponds on the gauge theory side correspond to the number of
fields entering the composite operator or equivalently the length
of the spin chain. The transition from the spin chains to the
sigma model relies on the coherent states
formalism~\cite{kruczenski03}.

Let $|ss\rangle$ be the state with the total spin $s$ and the projection of the
$z-$axis $S_z=s$. The coherent state for the spin$-s$ representation of the
$SU(2)$ group is defined as
\begin{equation}
|\vec{n}\rangle=\e^{iS_x \phi}\e^{i S_y\theta}|ss\rangle
\end{equation}
where $\vec{n}$ is the unit vector, $\vec n^2=1$,
\begin{equation}
\vec{n}=(\sin\theta \cos\phi,\sin\theta \sin\phi, \cos\theta)
\end{equation}
with $\theta$ and $\phi$ being spherical angles. Expanding the Hamiltonian of the
spin chain
\beq
H=\lambda/(4\pi^2) \sum_{k=1}^J (1/4- \vec S_k\cdot\vec S_{k+1})
\eeq
over the coherent states, one rewrites the partition function $\tr \e^{-Ht}$ in the standard manner as
a path integral over $\vec S_k=s \vec n_k$ with the following action
\begin{equation}
S(\vec{n})=s\sum_{k=1}^J\int dt\int _{0}^{1} d\tau\,
\vec{n}_k(\partial_t \vec{n}_k \times
\partial_{\tau}\vec{n}_k) - \frac{\lambda}{8\pi^2} s^2 \int dt
\sum_{k=1}^J(\vec{n}_k-\vec{n}_{k+1})^2\,,
\end{equation}
with $\vec n_{J+1}=\vec n_1$. In the long wave limit the vectors $\vec n_k(t)$
vary smoothly along the spin chain and, therefore, they can be approximated by a
function $\vec n(\sigma,t)$ with continuous $\sigma$ running between $0$ and the
chain length $J$ leading to
\begin{equation}
S= -s\int dt d\sigma\, \partial_t \phi \cos\theta  -
\frac{\lambda}{8\pi^2} s^2 \int dt d\sigma \left[(\partial_{\sigma}\theta)^2 +
(\partial_{\sigma}\phi)^2\sin^2\theta \right]
\label{action}
\end{equation}
It turns out~\cite{kruczenski03} that for $s=1/2$ this expression coincides with
the stringy action
\be
S_{\rm str}=\frac{R^2}{4\pi \alpha'}\int d\sigma d\tau \left[
G_{\mu\nu} \partial_\tau X^\mu \partial_\tau X^\nu-G_{\mu\nu} \partial_\sigma
X^\mu
\partial_\sigma X^\nu\right]
\ee
evaluated for the classical
string propagating in the background $ds^2 =G_{\mu\nu} dX^\mu dX^\nu$
\begin{equation}
ds^2= -dt^2 +d\psi^2 +d\varphi_1^2 +d\varphi_2^2
+2\cos(2\psi)d\varphi_1d\varphi_2\,.
\end{equation}
To see this, one fixes the gauge $t=\chi\tau$, takes the limit $\partial_\tau
X^i\to 0$ and $\chi\to \infty$ with $\chi\partial_\tau X^i= \rm fixed$ and
identifies the variables as
\begin{equation}
\varphi_2=-\frac12\phi, \qquad \psi=\frac12\theta\,.
\end{equation}
Then, one excludes $\varphi_1$ with a help of the classical equations of motion
and arrives at (\ref{action}).

The derivation of the effective action can be also generalized to the $SU(3)$
case~\cite{lopes} and to the string carrying both large Lorentz spin $S$ and the
$R-$charge $J$~\cite{ts}. One can improve the effective sigma model action
(\ref{action}) by calculating corrections involving higher derivatives of fields.

\subsection{Semiclassical string motion and integrable models}

We have argued above that the effective action for long wave excitations in the
compact spin chain coincides with the classical action of the sigma model on the
curved background relevant for calculation of the anomalous dimensions of the BMN
like operators. As the next step, one has to compare the corresponding solutions
to the equations of motion. To this end, one considers the bosonic part of the
superstring action on the $AdS_5\times S^5$ background. It is given by the sum of
two coset sigma models
\begin{equation}
S=\frac{\sqrt {\lambda}}{4\pi}\int d\sigma d\tau [G_{mn}^{AdS}\partial
y_{m}
\partial y_{n} + G_{kl}^{S^5}
\partial x^{k}\partial x^{l}]\,,
\end{equation}
where the string tension is proportional to the t'Hooft coupling.
 It is convenient to rewrite the action with constraint
imposed with the Lagrangian multiplier
\begin{equation}
S=\frac{\sqrt {\lambda}}{4\pi}\int d\sigma d\tau [\partial X_{m}
\partial X_{m}+ \Lambda_{x}(X^2 -1)+
\partial Y^{k}\partial Y^{k}+ \Lambda_{y}(Y^2 +1)]
\label{S-Polyakov}
\end{equation}
where $X_n$ $(n=1,\dots6)$ and $Y_k$ $(k=0,\dots,5)$ are the two sets of the
embedded coordinates in the flat $R^6$ space with the signatures $(6,0)$ and
$(4,2)$, respectively. The action has to be supplemented by the Virasoro
constraint for vanishing the two-dimensional energy momentum tensor
\be
 \dot{Y_k}\dot{Y_l} + Y_{k}' Y_{l}' + \dot{X_n}\dot{X_n}+ X_{n}'X_{n}'
=\dot{Y_k}Y_{k}' +\dot{X_n}X_{n}'=0
\label{Virasoro-const}
\ee
and by the periodic boundary conditions 
\begin{equation}
Y_k(\sigma +2\pi)=Y_k(\sigma), \qquad X_n(\sigma +2\pi)=X_n(\sigma)\,.
\label{boundary-cond}
\end{equation}
Due to the $SO(2,4)$ and $SO(6)$ symmetries, the classical action possesses the
set of the conserved charges
\begin{eqnarray}
S_{kl} &=& \sqrt{\lambda}\int d\sigma(Y_k\dot{Y_l}- Y_l\dot{Y_k})
\, ,
\nonumber \\
J_{nm} &=& \sqrt{\lambda}\int d\sigma(X_n\dot{X_m}- X_m\dot{X_l})\,.
\end{eqnarray}
Among them one distinguishes 6 Cartan generators: the energy $E=S_{05}$, the
Lorentz spins $S_{12},S_{34}$ and the $S^5$ angular momenta
$J_{12},J_{34},J_{56}$. These conserved charges parameterize general solutions to
the classical equations of motion~\cite{ft1}.

To describe a particular operator on the gauge theory side we have to identify
the corresponding solution to the classical equations of motion in the sigma
model~\re{S-Polyakov} subject to the constraints \re{Virasoro-const} and
\re{boundary-cond}. The simplest anzatz corresponding to a string located at the
center of the $AdS_5$ and rotating in the $S^5$ looks as follows
\begin{equation}
Y_5+i Y_0=\e^{it}\,,\qquad
X_{2i-1}+iX_{2i}=r_{i}(\sigma)\,\e^{i\omega_{i}\tau+i\alpha_i(\sigma)}\,.
\end{equation}
with $i=1,2,3$ and remaining $Y-$coordinates set to zero. Its substitution into
the sigma model action \re{S-Polyakov} yields the Lagrangian
\cite{Arutyunov:2003za}
\begin{equation}
L = \sum_{i=1}^{3} \left( r_{i}^{\prime 2} + r_{i}^2\alpha_{i}^{\prime
2} - \omega_{i}^2 r_{i}^2 \right) - \Lambda_x \sum_{i=1}^{3} \left( r_{i}^{2} - 1
\right) \, .
\label{system}
\end{equation}
Solving the equations of motion for $\alpha_i$ one gets $\alpha_{i}^\prime =
{v_i}/{r_{i}^2}$ with $v_i$ being the integration constants. The resulting
Lagrangian describes the integrable Neumann-Rosochatius system.%
It admits five independent integrals of motion: $v_1, v_2,
v_3$ plus two additional integrals which look as
\begin{equation}
I_{i} = r_{i}^2 + \sum_{j\neq i}^{3} \frac{1}{\omega_{i}^2 - \omega_{j}^2} \left[
\left( r_i r_{j}^\prime - r_i r_{j}^\prime \right)^2 + \frac{v_i^2
r_{j}^2}{r_i^2} + \frac{v_j^2 r_{i}^2}{r_j^2} \right]
\end{equation}
subject to $\sum_{i=1}^3 I_i=0$. The periodicity condition trades $v_i$ for three
integers $m_i$ and $I_i$ for two integers $n_i$. As a result, the energy depends
on the frequencies $\omega_i$ and five integers. These variables are not
independent since Virasoro constraint imposes a relation between them. For this
type of string motion, the infinite set of conserved charges in the sigma model
is parameterized by a finite set of integrals of motion \cite{am1,am2}. The
energy corresponding to classical solutions to the string sigma model defines the
anomalous dimension of the dual composite scalar operators in the ${N}=4$
SYM theory. There are many examples of such correspondence discussed in the
literature, initiated in Ref.~\cite{ft1} and reviewed in \cite{tsey1}

\subsection{Generic spin chains/string correspondence}

We have presented above the examples of the derivation of the
string from the finite dimensional integrable model and vice versa. Turn now to the
discussion on  generic
features of their interrelation. There are several questions
which require immediate answers. First question concerns the
dictionary between type of the spin chain and the corresponding
solution to the stringy equation of motion. Generic motion of the
string in $AdS_5\times S_5$ is characterized by five quantum
numbers $S_1,S_2,J_1,J_2,J_3$ where $S_1,S_2$ correspond to the
values of Casimirs of the Lorentz group, while $J_i$ correspond to
charges with respect to R symmetry on the gauge theory side.
Solutions to the stringy equation of motion are mapped into the
operators on the gauge theory side with the same quantum numbers
and therefore the type of the spin chain governing the
renormalizaton of these operators is fixed by their quantum
numbers as well. For instance operators with two nonvanishing
R-charges $J_1,J_2$ are described by SU(2) spin chains, while
operators with single Lorentz spin S  by SL(2,R) spin chain. Most
general operators are governed by $SO(4,2|2)\times SO(6)$ superspin chain.

Since the anomalous dimension is identified with the string energy
evaluated on the particular solution to the equation of motion it
depends on the gauge coupling constant through its tension.
Generically this dependence is quite complicated but to compare
this stringy prediction with the one-loop gauge theory calculation
one has to expand the all-loop prediction in  powers of the
coupling constant. In the simplest case of the  large quantum numbers
energy depends on the coupling constant analytically. Hence one
can derive one-loop energy  from the
expansion for solution with $(J_1,J_2)$ quantum numbers
follows \cite{tsey1}
\beq E_{str}= \frac{2}{\pi^2} K(x)[E(x)-(1-x)K(x)],\qquad
\frac{J_2}{J_1+J_2}=1 - \frac{E(x)}{K(x)}
\eeq
where $K(x)$ and $E(x)$ are the standard elliptic integrals
of the first and second kind correspondingly.
This rather complicated expression for
the anomalous dimensions of the operators of the type
$Tr\Phi_1^{J_1}\Phi_2^{J_2}$ exactly coincides
with the one loop anomalous
dimensions derived from the corresponding SU(2) spin chain. This
is very nontrivial check of the relation between spin chains and
string dynamics. Similar one loop coincidence has been found for
operators with more complicated set of quantum numbers moreover
there is precise coincidence for higher integrals of motion as well.

The agreement at one-loop level is perfect therefore the natural
next step is the  examination of higher loops in N=4 YM theory. It
turns out that integrability holds at two loops and for instance
in SU(2) sector the renormalization of the operators at two loops
is governed by the integrable spin chains with next-to nearest neighbor
spin chain \cite{BeiKriSta03}
\beq
H_{2 loop}= \frac{\lambda}{8\pi^2}\sum_{k=1}^{J}( 1- P_{k,k+1}) +
\frac{\lambda^2}{128\pi^4}\sum_{k=1}^{J}(-4 +6P_{k,k+1} - P_{k,k+1}P_{k+1,k+2}
-P_{k+1,k+2}P_{k,k+1}) + O(\lambda^3)
\eeq
where $P_{i,j}$ is the permutation operator of i-th and j-th sites.
For the $S=1/2$ it can be represented as  $(\vec{S_i} \vec{S_j}
-const)$.

At three loops there is some
disagreement at the subleading in 1/J terms which implies some
additional complications \cite{callan}. There are some attempts to get the
all-loop results using some generalization of the Bethe anzatz
approach which includes the essential stringy S-matrix \cite{staudall}. This
approach fits perfectly with the recent 3-loop results in YM
theory in noncompact sector \cite{lipatov3loop} however its status as the correct all loop answer is
unclear.

\section{N=4 SYM and quantum string; pp-wave limit}
At present, explicit quantum solution to the string theory in the
$AdS_5\times S^5$ background is not available. This makes it
impossible to compare the string spectrum with the complete set of
the operators in ${N}=4$ SYM theory. Hopefully the
explicit solution to the string theory can be found~\cite{metzaev}
in the case of limiting geometry defined as the Penrose limit of
the $AdS_5\times S^5$ \cite{bmn,Beisert:2002bb,Gross}. The Penrose
limit can be described as the region around the null geodesic in
the $AdS_5 \times S^5$. Introducing new variables
\begin{equation}
x^{+}= \frac{t+\chi}{2\mu},\quad x^{-}=\mu R^2 (t-\chi)
\end{equation}
with $\chi$ being an angular variable in $S^5$ and $\mu$ being
some scale, one takes the limit $R \rightarrow \infty$ and
recovers the pp-wave metric
\begin{equation}
ds^2= -4dx^{+}dx^{-} - z^2 dx^{+2} + \sum_{i=1}^{8}dz_i^2\,.
\end{equation}
Here eight flat transverse coordinates $z_i$ come both from the
$S^5$ and $AdS_5$ parts of the geometry. In this metric, the
string behaves as a particle rotating with large angular momentum
$J$ along $\chi$ angular coordinate in $S^5$. The light-cone energy
of the string reads
\begin{equation}
H=2p^{-}=i(\partial_{t}+\partial_{\chi})=(\Delta -J)\,.
\end{equation}
For $R\to \infty$ it takes finite values provided that the
following double scaling limit is considered
\begin{equation}
R \rightarrow \infty,\qquad \Delta\sim J\rightarrow \infty, \qquad
\frac{J^2}{R^4} = \mbox{const}.
\end{equation}

Quantization of the string propagating in this background reduces
to the quantization of the oscillators. As a result, the exact
spectrum of the type IIB string in the pp-wave background looks as
\begin{equation}
\label{dim} \Delta-J=\sum_k N_k\sqrt{1+\frac{\lambda k^2}{J^2}}\,,
\end{equation}
where $k$ labels the Fourier modes, $N_k$ denotes the total
occupation number of oscillatory mode and the condition
$P=\sum_{k}kN_k=0$ is imposed.

On the gauge theory side (\ref{dim}) defines anomalous dimension
of certain Wilson operators in the ${N}=4$ SYM theory. The
length of the string $J$ equals the number of the constituents of
the composite operator. The ground state of the string can be
identified with the operator built from scalars $Z=\Phi_1
+i\Phi_2$
\begin{equation}
|0,J \rangle \leftrightarrow {\rm tr}\,Z^J\,.
\end{equation}
These operators have the charge $J$ with respect to ``rotation
plane" in pp-wave. The oscillatory excitations of the string
ground state correspond to the insertions of other scalar fields.
For example, the so-called BMN operators in the ${N}=4$
SYM theory can be mapped into the stringy modes as follows
\ba
 a_{0}^{i+} |0, J \rangle &\Leftrightarrow& {\rm tr}\,\Phi_i Z^J
\nonumber
\\
 a_{n}^{i+}a_{-n}^{j+} |0, J \rangle &\Leftrightarrow& \sum_{l}e^{2\pi i n l/J}
{\rm tr}\,\Phi_iZ^{l}\Phi_{j}Z^{J-l}
\ea
One can deduce from these
expressions that calculation of the stringy spectrum corresponds
to diagonalization of the mixing matrix for Wilson operators on
the gauge theory side.

The energy of the string in the pp-wave limit is a function of the
ratio of the coupling constant and angular momentum,
${\lambda}/{J^2}$. It is expected that expansion of this function
in powers of ${\lambda}/{J^2}$ should reproduce perturbative
series for the anomalous dimension of the corresponding Wilson
operators in the weak coupling regime.
%
%
On the other hand, the one-loop ${N}=4$ dilatation
operator in the sector of scalar operators coincides with the
Hamiltonian of the Heisenberg $SO(6)$ spin chain \cite{mz}. This
allows one to map stringy states into the spin chain states. The
correspondence is very precise for Wilson operators built only
from two complex scalars in which case the $SO(6)$ spin chain
reduces to the conventional Heisenberg $SU(2)$ spin$-1/2$ chain.
Then, the ground state in the string theory corresponds to all
spins aligned in the same direction in the isotopic space while
the stringy excitations correspond to the flip of some spins along
the chain.

\section{Strong coupling phenomena in nonconformal case; N=2 SYM}

In this section we shall discuss the dual supergravity description of $N=2$
SYM theory. First, we shall describe the background involved into
consideration and compare it with $N=4$ case. Then we discuss a few
field theory phenomena which can be easily captured at the
supergravity approximation. To start with remind the generic
features of $N=2$ theory. The pure gauge theory contains $N=2$
supermultiplet in the adjoint representation of the gauge group
which involves vector field, two Majorana spinors and complex
scalar $\Phi$. Theory is asymptotically free and  the $\beta$ function
gets contribution only from one loop.
The Lagrangian  classically has $SU(2)\times U(1)$ global $R$
symmetry group but the $U(1)$ part of R symmetry group is broken at the
quantum level down to $Z_{4N}$. Theory has vacuum valley
parameterized by the vacuum expectation values of the scalar field
and the nonperturbative metric on the moduli space of the vacuum
manifold has been found in \cite{sw}.

The metric in the dual supergravity description reads as \cite{polchinski20}

\beq
ds^2=H^{-1/2}\eta_{\mu\nu}dx^{\mu}dx^{\nu} + H^{1/2} [d\rho^2
+\rho^2d\theta^2 +\delta_{mn}dx^ndx^m]
\eeq
and for  higher form field  involved in the solution

\beq \tilde{F_5}
=d(H^{-1}dx^0\wedge \dots dx^3) +*d(H^{-1}dx^0\wedge \dots dx^3)
\eeq

\beq
c+ib=4\pi\alpha^{'}g_sNlog\frac{z}{\rho_0}
\eeq
where $\rho^2=(x_4^2 +x_5^2), z=\rho e^{i\theta}, r^2=x_6^2 +\dots+x_9^2$ and H is
known function of the radial coordinates. The solution involves
the complex scalar field $c+ib$ and form fields in NS-NS and R-R sectors.

\beq
\tilde{F_5}=F_5 -C_2\wedge H_3, \qquad H_3=dB_2
\eeq
The supergravity solution describes  a bound states of $N$
fractional $D3$ branes  on the orbifold $C^2/Z_2$ and the gauge theory
can be thought of as theory on their common worldvolume.

Let us turn to the description of a few quantitative phenomena.
The parameters of YM theory are derived from the Born-Infeld action
on the $D3$ brane worldvolume in the low
energy limit  on the gravity solution

\beq
\label{con}
\frac{1}{g_{ym}^2}=\frac{1}{16\pi^2\alpha^{'}g_s}\int B_2
=\frac{N}{4\pi^2}log\frac{\rho}{\rho_0} \eeq and

\beq
\theta_{YM}=\frac{1}{2\pi^2\alpha^{'}g_s}\int C_2=-2N\theta \eeq
The supergravity solution enjoys the  symmetry
\beq
\theta \to \theta+\frac{\pi k}{2N}
\eeq
and using the relation with
$\theta_{YM}$ one immediately recognizes this symmetry as the
counterpart of $Z_{4N}$ symmetry on the gauge theory side. To get
$\beta$ function from the supergravity solution it is necessary to
make the proper identification of the scale $\mu$ on the gauge
side in terms of the gravity variable $\rho$. To this aim it is
convenient to consider the protected operator and look at its
scale properties. The complex scalar is the convenient object to
deal with and using its gravity identification
$\phi=(2\pi\alpha')^{-1}z$ the following relation can be easily
found
\beq
\rho=2\pi \alpha'\mu
\eeq
Inserting this relation
into gravity solution (\ref{con}) we immediately derive
\beq
\frac{1}{g_{YM}^2}=\frac{Nlog(\mu/\Lambda)}{4\pi^2}
\eeq
that is the
correct $\beta$-function of $N=2$ theory. Note that the full
non-perturbative effective action for the gauge theory has not
been derived yet on the gravity side. The problem with its derivation is due  to the
so-called enhancon phenomenon \cite{polchinski20}. It corresponds to the singularity
of the function H involved into the solution which can be
resolved only upon additional stringy modes are taken into
account.

\section{Strong coupling phenomena in nonconformal case; $N=1$ SYM}
The $N=2$ supersymmetric theory is essentially different from
realistic models by virtue of its having an infinite number
of vacuum states.
More interesting phenomena are known to exist in $N=1$ SYM theory. First, let
us remind the general features of $N=1$ theory and emphasize the
differences with $N=2$ case. The $N=1$ theory contains vector
supermultiplet involving vector field and Majorana gluino in
adjoint representation. The action reads as

\beq
L_{N=1}=\frac{1}{g_{YM}^2}\int d^4 xtr(F^2 +i\bar{\lambda}D\lambda)
\eeq
and the theory has many similarities with QCD, in particular it is
asymptotically free and develops the mass gap. Contrary to $N=2$
case theory has finite number of the vacuum states, namely $N$
states for $SU(N)$ gauge group. There are  anomalies in the
theory which belong to the single anomaly multiplet. One anomaly
current concerns the trace of the energy stress -tensor generating
dilatations and second current involves R-symmetry phase rotation
of gluino field. Both symmetries are broken at one loop. Contrary
to $N=2$ case where $\beta$-function has only one loop contribution
in N=1 case all loops matter and exact perturbative result can be found
\cite{nsvz}

\beq
\beta=-\frac{Ng_{YM}^2}{16\pi^2}(1-\frac{Ng_{YM}^2}{8\pi^2})^{-1}
\eeq
All contributions beyond one-loop involve IR region therefore
Wilsonian $\beta$ function is one-loop exact.

Let us explain the fate of U(1) R-symmetry. As we have already
mentioned U(1) is broken down to $Z_{2N}$ by the one-loop anomaly.
Moreover it is further broken spontaneously down to $Z_2$ by the
gluino condensate developed due to the nonperturbative effects

\beq <tr\lambda^2>= \Lambda^3 e^{2\pi ik/N},\qquad k=0,1,\dots,N-1
\label{cond} \eeq
where $\Lambda$ is the scale generated by the dimensional
transmutation phenomena. The gluino condensate is the order
parameter in the theory and k in (\ref{cond}) labels k-th vacuum
state.

\subsection{Supergravity solution}

Turn now to the description of the supergravity solution for N=1
theory. Actually there are two most popular geometries responsible
for this case; Maldacena-Nunez background \cite{mn} and Klebanov-Strassler
background \cite{ks} which are related to each other.
In what follows we shall focus on Maldacena-Nunez
pattern corresponding to the wrapped branes. Earlier we discussed
D3 branes embedded into ten dimensional space however to get dual
to $N=1$ theory  the stack of $N D5$ branes wrapped around
two-circle in ten dimensions has to be added. The gauge theory
emerges on the worldvolume of wrapped $D5$ branes in the
following background

\beq ds^2=e^{\Phi}dx^2 +e^{\Phi}g_sN[e^{2h}(d\theta_1^2 +
sin^2\theta_1 d\phi_1^2) +d\rho^2 +\sum_{a=1}^3(\omega^a - A^a)^2]
\eeq

\beq e^{2\Phi}=\frac{sinh\rho}{2e^{h}}\eeq

\beq F_3=2g_sN\prod_{a=1}^{3}(\omega^a-A^a)
-g_sN\sum_{a=1}^{3} F^a\wedge\omega^a \eeq

where \beq A^1=-1/2a(\rho)d\theta_1,\qquad A^2=1/2a(\rho)
sin\theta_1d\phi_1, \quad A^3=-1/2cos\theta_1d\phi_1 \eeq

\beq e^h=\rho coth2\rho -\frac{\rho^2}{sinh^22\rho}-1/4,\quad
a(\rho)=\frac{2\rho}{sinh\rho} \eeq
The left invariant forms read as
\be
2\omega^1=cos\psi d\theta_1+sin\psi sin\theta_2 d\phi_2,
\quad 2\omega_2=sin\psi d\theta_2 -cos\psi sin\theta_2 d\phi_2,
\quad
2\omega^3=d\psi+cos\theta_2d\phi_3
\ee
and $F^a=\nabla A^a$.
The solution is nonsingular and  involves five angles and the
radial coordinate $\rho$ which we shall relate with the energy
scale in the gauge theory. The parameters of the gauge theory are
identified upon the substitution of the supergravity solution into
the low energy expansion of Born-Infeld action

\beq \frac{1}{g_{YM}^2}=\frac{1}{16\pi^3 g_s}\int_{S^2}e^{-\Phi}
\sqrt{detG}=\frac{N}{4\pi^2}\rho tanh\rho \eeq

\beq
\theta_{YM}= - \frac{1}{2\pi g_s}\int _{S^2}C_2 =-N\psi
\eeq
It is clear from the solution that large $\rho$ corresponds to the
UV region of gauge theory where coupling constant is small and
U(1) R-symmetry rotation corresponds to shift in angular variable
$\psi$.

\subsection{Physics of N=1 SYM from supergravity solution}

We now move to the descriptions of the features of N=1 theory described
above in terms of the supergravity solution. Let us first consider the
pattern of  U(1) R-symmetry breaking. Since in UV region we
expect breaking of U(1) down to $Z_{2N}$ the same has to be true
on the gravity side. To this aim let us consider large $\rho$
limit when $a(\rho)\to 0$. In this case rotation of $\psi$ is
the isometry of the metric and noninvariance is due to $C_2$ flux
only. The immediate check shows that

\beq \psi\to \psi +\frac{2\pi k}{N}
\eeq
is the symmetry of the solution which precisely corresponds to the
$Z_{2N}$ symmetry on the gravity side.

To derive the additional spontaneous breaking down to $Z_2$ let us note
that the scale factor $a(\rho)$ is multiplied by the $cos\psi$ or
$\sin \psi$ which implies that at arbitrary $\rho$ only
\beq
\psi \to \psi + 2\pi N
\eeq
is the symmetry of solution which
precisely matches  the field theory result.

To obtain the relation between the radial variable $\rho$ and the
energy scale $\mu$ it is convenient to use the identification

\beq
<\lambda^2>\leftrightarrow a(\rho)
\eeq
which implies the relation

\beq \frac{\Lambda^3}{\mu^3}=\frac{2\rho}{sinh2\rho}\eeq
Note that this identification involves the dimension of the
protected operator which is not influenced by the quantum
corrections. With such identification exact $\beta$ function can
be calculated  using the relation

\beq \beta_{YM}=\frac{\partial g_{YM}}{\partial \rho}
\frac{\partial \rho}{\partial(log\mu/\Lambda)} \eeq
which yields at large $\rho$

\beq \frac{\partial g_{YM}}{\partial
\rho}=-\frac{Ng_{YM}^2}{8\pi^2}\eeq

\beq \frac{\partial
\rho}{\partial(log\mu/\Lambda)}=3/2(1-\frac{Ng_{YM}^2}{8\pi^2})^{-1}
\eeq
Combination of two expressions amounts precisely to the
exact perturbative $\beta$-function \cite{divec}. The gravity solution predicts
also the nonperturbative corrections to the $\beta$-function
however their meaning on the gauge theory side is not clarified yet.

\section {Yang-Mills theory without supersymmetry and anomalous
dimensions}

In this section we shall present the examples  how duality works in the
pure gauge theory without SUSY at all. On the first glance
nothing can be said since we departure the safe supersymmetric
world however it turns out that it is not the case. We shall
discuss here only two issues concerning well established
results avoiding  more speculative claims distributed in
the literature. First, we shall mention the integrable
structure of the one-loop dilatation operator  in the selfdual
gluonic sector and its stringy realization. Secondly, we shall
explain the predictions for the anomalous dimensions of some
YM operators at the strong coupling regime which are universal
for the gauge theories.

\subsection{Classical string and gluonic operators}

A particular case of certain  local multi-gluon higher-dimension
operators of the type
\begin{equation}
\label{LocalGluonOper}
\prod_{j = 1}^L F_{\mu_j \nu_j} (0)
\end{equation}
in pure gluodynamics have been addressed recently and found to possess integrable
structures corresponding to the compact spin-one Heisenberg magnet
\cite{Ferretti:2004ba}. Without loss of generality we can rephrase their analysis
entirely in Euclidean space. The strength tensor can be decomposed into
irreducible components
\begin{equation}
F_{\mu\nu} = \eta^A_{\mu\nu} F^A_{+} + \bar\eta^A_{\mu\nu} F^A_-
\end{equation}
with the help of 't Hooft symbols, $O (4) \sim SU (2) \otimes SU(2)$.
The selfdual and anti-selfdual components transform as $(1, 0)$ and
$(0, 1)$, respectively. The part of the RG Hamiltonian responsible for
eigenvalues of autonomous components does not change the number of
fields in the local gluonic operator (\ref{LocalGluonOper}). By matching
the coefficient of different irreducible components, extracted by means
of projectors ${P}^P_{(j_1, j_2)}$ for spin-$j$ and parity $P$,
to available one-loop calculations gluonic operators up to dimension eight,
the pair-wise Hamiltonian was found to be  \cite{Ferretti:2004ba}
\begin{equation}
{H}_{12}
= 7
\left( {P}_{(2,0)} + {P}_{(0,2)} \right)
+
{P}_{(1,0)} + {P}_{(0,1)}
- 11
\left( {P}_{(0,0)}^+ + {P}_{(0,0)}^- \right)
+ 3
{P}_{(1,1)}^-
\, .
\end{equation}
The projection on the selfdual operators, i.e., built from products
of $F^A_+$, reduces the above Hamiltonian to
\begin{equation}
{H}_{12}^{\rm sd}
= 7
{P}_{(2,0)}
+
{P}_{(1,0)}
- 11
{P}_{(0,0)}
\, ,
\end{equation}
where the projection operators extract maximal-spin, antisymmetric and trace
components and have the following obvious representation
\begin{eqnarray*}
{P}_{(2,0)}
F^A_+ F^B_+
&=&
\ft12
\left(
F^A_+ F^B_+ + F^A_+ F^B_+ - \ft23 \delta^{AB} F^C_+ F^C_+
\right)
\, , \\
{P}_{(1,0)}
F^A_+ F^B_+
&=&
\ft12
\left(
F^A_+ F^B_+ - F^A_+ F^B_+
\right)
\, , \\
{P}_{(2,0)}
F^A_+ F^B_+
&=&
\ft13
\delta^{AB} F^C_+ F^C_+
\, .
\end{eqnarray*}
They can be easily related to the permutation $P F^A_+ F^B_+ = F^B_+
F^A_+$, trace $K F^A_+ F^B_+ = \delta^{AB} F^C_+ F^C_+$ and identity
$I F^A_+ F^B_+ = F^A_+ F^B_+$ operators,
$$
{P}_{(2,0)}
=
\ft12 \left( I + P \right) - \ft13 K
\, , \quad
{P}_{(1,0)}
=
\ft12 \left( I - P \right)
\, , \quad
{P}_{(0,0)}
=
\ft13 K
\, .
$$
Then, the pair-wise hamiltonian can be brought to the form
\begin{equation}
{H}_{12}^{\rm sd}
= 4 I_{12} + 3 P_{12} - 6 K_{12}
= 7 + 3 \bit{s}_1 \cdot \bit{s}_2 \left( 1 - \bit{s}_1 \cdot \bit{s}_2 \right)
\, ,
\end{equation}
where in the second equality we have used the representation in terms of
spin-one $SU (2)$ generators. This is a Hamiltonian of an exactly solvable
spin-one Heisenberg magnet which can be diagonalized by means of the
Bethe Ansatz \cite{KRS81,TTF83}.

The main feature of this class of operators is that their
anomalous dimensions scale as L. Recently the stringy dual of this class
of operators has been found \cite{tsey05}. It turns out that the
corresponding string motion involves rotation in two independent
planes in $AdS_5$ with quantum numbers (S,S,0,0,0). The explicit
expression for the stringy energy  reads as \cite{tsey05}
\beq
E=2\sqrt{mS} + O(S^{2/3})
\eeq
at small S, where m is the string winding number. This is equivalent
to the flat space result. However if we shall be interested
in the long string then $S>>1$ and energy goes as
\beq
E=2S +\frac{3}{4}(4m^2S)^{1/3} + \dots
\eeq
It turn out that the stringy solution is unstable at large S however
the energy is linear in S at the intermediate S in the stability
region which is in agreement with the one loop calculation in the
gauge theory.

Note also that the integrable structure involving anisotropic XXZ spin chain
has been also found in N=1 SYM theory \cite{dive}.

\subsection{Predictions for the anomalous dimensions at
strong coupling}
 In general string  moves
both in the $AdS_5$ and $S^5$ and could have large angular momenta in both
spaces. Contrary to the pure scalar operators, when the comparison with the loop
expansion on the gauge side can be performed for the operators with large $R$
charge, the situation with operators carrying large Lorentz spin $S$ is more
subtle. The folded closed string rotating in the $AdS_5$ yields the dependence
for the anomalous dimensions of twist-two operators $F_{+ \perp } (D_+)^S
F_{+\perp}$ at large coupling \cite{gkp}
\be
\gamma^{\rm (tw=2)}_S =  \frac{\sqrt{\lambda}}{2\pi} \ln S^2\,.
\label{twist-2-open}
\ee
This result can be generalized to higher twist operators of the form $F_{+\perp}
D_+^{S_1} F_{+\perp} \dots D_+^{S_{L-1}} F_{+\perp}$. The energy of the
corresponding revolving string coincides with the energy of the classical
Heisenberg spin chain of the length $L$ and leads to \cite{bgk}
\begin{equation}
\gamma^{{\rm (tw=}L)}_{S_1, S_2, \dots , S_{L-1}} = \frac{\sqrt{\lambda}}{2\pi}
\ln q_L(S_1,S_2,\dots , S_{L-1}) \,.
\label{twist-L-open}
\end{equation}
Here $q_L$ is the highest integral of motion of the spin chain. For $S_k\sim S
\gg 1$ with $k=1,\ldots, L-1$ one has $q_L \sim S^L$.

Notice that logarithmic scaling of the anomalous dimensions is a universal
feature of Wilson operators with large Lorentz spin in gauge theories, unrelated
to the presence of supersymmetry~\cite{Korchemsky88,km}. However stringy
description of this scaling at weak coupling remains unknown and it is doubtful
whether such classical string sigma model solutions exist.
One should expect instead that $\sim \ln S$ behavior at weak coupling is driven
by the quantum sigma model.

Integrability phenomenon offers the possibility to extend the gauge/string
duality beyond the special class of classical string solutions described above.
Namely, instead of comparing particular solutions one can identify the integrable
structures corresponding to \textsl{quantum} spin chains describing dilatation
operator on the gauge theory side and  to classical equations of motion on the
string theory side~\cite{kmmz}. It turns out that in both cases integrability is
encoded in the properties of Riemann surfaces.

For the quantum spin chains, the appearance of Riemann surfaces within the
framework of the Bethe Ansatz is not surprising. As we already mentioned
semiclassical solutions to the Baxter equation are determined by the
properties of the spectral curve whose genus is proportional to the number of
sites in the chain. In the case of the BMN like operators, the number of constituent scalar
fields goes to infinity in the thermodynamical limit and, therefore, the
corresponding Riemann surface would have an infinite genus. However choosing the
appropriate values of the integrals of motion it is possible to pinch almost all
handles and obtain a finite genus surface. It is this degenerate surface which
parameterizes general solutions to the classical equations of motion of the
string~\cite{kmmz}. The agreement between semiclassical solutions to
the Bethe Ansatz equations and solutions to the classical string equations of
motion has been carefully checked up to the two-loop level~\cite{kmmz}.

Although the correspondence between one- and two-loop dilatation operators in the
${N}=4$ SYM theory Yang-Mills theory, stringy states and integrable
quantum spin chains is well established, the situation with higher loops in the
perturbation theory is unclear. Several proposals have been made concerning
integrable structures behind higher loop dilatation operator~\cite{SS,bdm}. At
the same time, starting from three-loop order the discrepancy seems to arise
between expressions for the anomalous dimensions of composite scalar operators
with large-$R$ charge and energy spectrum of the string~\cite{callan}. More work
is needed to clarify this issue.

\subsection{Open string picture for anomalous dimensions}

There exists an alternative description of logarithmic growth of the anomalous
dimension of Wilson operators with large Lorentz spin, Eq.~\re{twist-2-open} and
\re{twist-L-open}, in terms of Wilson loops in the gauge theory and open strings
on the $AdS_5$ background. This picture relies on the correspondence between
anomalous dimension of the composite operators with large number of light-cone
derivatives and the so-called cusp anomaly of Wilson
loops~\cite{Korchemsky88,km}. It was shown a long time ago \cite{polyakov2} that
Wilson loop $W[C]=\tr\{P\exp(i g \int_C dx^\mu A_\mu(x))\}$ acquires a nontrivial
anomalous dimension $\Gamma_{\rm cusp}(\lambda, \theta)$ if the integration
contour $C$ has a cusp
\be
\vev{W[C]} \sim \mu^{\Gamma_{\rm cusp}(\lambda,\theta)}\,,
\ee
with $\mu$ being a UV cut-off. The cusp angle $\theta$ is restricted to the
interval $[0,2\pi[$ in Euclidean space but is unrestricted in Minkowski. The
correspondence between the anomalous dimension of twist-2 spin operators with
large Lorentz spin $S$ and the cusp anomaly looks as
follows~\cite{Korchemsky88,km}
\begin{equation}
\gamma^{\rm (tw=2)}_S(\lambda)=2\Gamma_{\rm cusp}(\lambda, \theta=\ln S)
\label{gamma=cusp}
\end{equation}
and it is valid for arbitrary coupling constant $\lambda$. At weak coupling and
$\theta\gg 1$ one has
\be
\Gamma_{\rm cusp}(\lambda, \theta)=\theta \left[\frac{\lambda}{4\pi^2} +
{O}( \lambda^2)  \right]
\ee
with perturbative coefficients known up to three-loop order. The
calculation of $\Gamma_{\rm cusp}(\lambda, \theta)$ at the strong coupling can be
effectively done via the open string picture. In this limit the Wilson loop is
proportional to the area of the minimal surface swept by an open string which
penetrates into the fifth AdS dimension and whose ends trace the integration
contour $C$ in Minkowski space. This leads to~\cite{kru2,mak}
\be
\Gamma_{\rm cusp}(\lambda, \theta)=\theta
\left[\left(\frac{\lambda}{4\pi^2}\right)^{1/2} + {O}(\lambda^0) \right]
\label{cusp-strong}
\ee
for $\theta\gg 1$. Being combined together, Eqs.~\re{gamma=cusp} and
\re{cusp-strong} reproduce the strong coupling result \re{twist-2-open} based on
the folded closed string picture~\cite{gkp}.

The correspondence \re{gamma=cusp} can be generalized to higher twist operators.
In that case, the anomalous dimension of the Wilson operator built from $L$
constituent fields and the total number of derivatives $S$ such that $S\gg L$ can
be mapped into anomalous dimension of the product of $L$ Wilson loops in the
fundamental representation of the $SU(N_c)$ group and the total number of cusps
varying between $4$ and $2L$~\cite{bgk}. At large $N_c$, the expectation value of
the product of Wilson loops factorizes into the product of their expectation
values. This implies that at strong coupling the area of the minimal surface
corresponding to the product of $k=2,\ldots,L$ Wilson loops with cusps is given
by the sum of $k$ elementary areas leading to
\be
2 \,\Gamma_{\rm cusp}(\lambda, \theta=\ln S) \le \gamma^{{\rm (tw=}L)}_S(\lambda)
\le L\, \Gamma_{\rm cusp}(\lambda, \theta=\ln S)\,.
\label{band1}
\ee
We remind that the anomalous dimensions of higher twist operators are not solely
determined by the total number of derivatives $S$. They form instead a band whose
internal structure at weak coupling is governed by integrals of motion of the
quantum Heisenberg $SL(2)$ magnet. Eq.~\re{band1} defines the boundaries of the
band both at strong and weak coupling.

\section{Conclusion}
In this brief review we have tried to cover  the results obtained
and the recent directions of the development of the issues
concerning the duality between the gauge theories with or
without supersymmetry and closed string theory in the curved
background. It is clear that only the first steps have been
made but even these restricted progress seems to prove
that the approach is very promising. After being
rather academic issue during   decades
gauge/string duality becomes powerful approach to analyze
gauge theory both at weak and strong coupling regimes. We would
like to emphasize that this approach already yields some
predictions on the gauge theory side which have been verified.
On the other hand the duality has inverse impact on the
string theory  since it involves closed string and potentially
could be useful to get new insights in the quantum gravity.

Of course the most interesting theory is the Standard Model
but duality does not provide much information about it yet.
However it is clear that approach should work in this case as well
and the search for the corresponding gravity background is
in progress now. Such gravity background potentially
could be responsible for the explanation of the confinement
phenomena in QCD at the strong coupling. In the weak coupling
high energy regime the great hopes are related with the possible
resummation of the perturbation series. In spite of the absence
of the cancellations familiar in N=4 case in theories
without SUSY some universal features of the perturbative
series are expected to be captured by the string theory.
The possible stringy realization
of the Regge regime in QCD is especially interesting since it could provide the
clear interpretation of the effective degrees of freedom
in this kinematics. Some first steps in this direction
have been done in \cite{gkk,janik,ps}.

The crucial role in the future developments will be played by
the hidden integrability of the theory on the both sides
of the correspondence. It just reflects the hidden symmetries
which have been missed in the previous studies.
In all
cases that we have mentioned, integrability appears as a hidden symmetry of an
underlying effective theory. It  describes elementary fields living on the
light cone in case of renormalization group evolution in QCD and
more generic operators in supersymmetric extensions.

A number of nontrivial questions still have to
be answered. The most obvious and at the same time the most profound and most
hard question is ``What is the origin of integrability?'' or in other words
``What is the symmetry, if any, of the gauge theory which leads to it?''
The first steps in the clarification of the hidden symmetries just in the
Lagrangian of YM theory have been done in \cite{wadia,bena,dolan,alday}
More specific questions  concern
issues like the fate of the integrability of the dilatation operator in SYM at
higher orders of perturbation theory, the integrability of the SYM dual sigma
models on curved backgrounds, matching of integrable structures on both sides of
the correspondence, just to name a few. Ultimately, if the integrability is
indeed the property of the full quantum Yang-Mills theory as well as the dual
string theory, it will provide the most sophisticated test of duality and endow
us with a powerful machinery to tackle the strong-coupling regime of field
theories.

In conclusion we mention certain results obtained recently. For example, it
was shown how the U(1) problem \cite{u1} is solved in the dual gravity
theory and the metric for nonconformal supersymmetric gauge theory with
fundamental matter was found \cite{flavour}. Some authors undertook
to obtain the physical characteristics of mesons in the
standard QCD in the dual theory
\cite{qcd}.

We barely touched on the problem of the dual string description
from the first principles. Only minimal progress has been made thus far
toward resolving this issue. Nevertheless, a few promising results
deserved to be mentioned. A new approach to the summation
of instanton effects has been developed \cite{nekrasov} which has provided
a basis for the hypothesis that the gauge theory actually
plays the role of an effective theory of microscopic gravitational
degrees of freedom \cite{iqbal}.

On the other hand it was noticed \cite{gopakumar} that loop calculations in the
four-dimensional field theory may be reformulated as tree diagrams in
five-dimensional
space in
$AdS_5$ metric.
Finally a new mechanism of generating an effective gravity theory
from the "condensation" of special states in gauge theory was proposed
in \cite{lunin}.
At the same time the key problem of the physical mechanism underlying
the generation of the metric condensate in quantum gravity
remains to be solved despite some positive trends.

We are grateful for A.Belitsky, V.Braun, G. Korchemsky for the collaboration
on the issues considered in the review. We also thanks A. Gerasimov,
A.Tseytlin, K. Zarembo, A.Marshakov, A.Mironov, A.Morozov, N. Nekrasov,
Yu. Makeenko
for sharing their insights on this subject. The work has been supported
in part by  CRDF grant RUP2-261-MO-04,
and RFBR grant 04-011-00646.


\end{document}